\documentclass[11pt]{article}

\usepackage[final]{acl}

\usepackage{times}
\usepackage{latexsym}
\usepackage{subcaption}
\usepackage{multirow}
\usepackage{multicol}
\usepackage{balance}
\usepackage{enumitem}

\usepackage[T1]{fontenc}

\usepackage[utf8]{inputenc}

\usepackage{microtype}

\usepackage{inconsolata}

\usepackage{graphicx}

\title{Beyond Semantic IDs: Encoding Business-Value Ranking into Document Identifiers for Generative Retrieval}

\author{
    \textbf{Gui Ling, Zhihong Chen, Yu Li, Tong Xiong, Kunhai Lin}, \\
    \textbf{Kaixuan Zhang, Yuliang Yan\thanks{Corresponding author.}, Dan Ou, Haihong Tang, Bo Zheng} \\
    Taobao \& Tmall Group of Alibaba \\
    \texttt{\{\ linggui.lg, jhon.czh, ly328242, xiongtong.xt, linkunhai.lkh, } \\
    \texttt{zhangkaixuan.zkx, yuliang.yyl, oudan.od, piaoxue, bozheng\ \}@taobao.com}
}

\begin{document}
\maketitle

\begin{abstract}
Generative Retrieval (GR) formulates retrieval as a sequence-to-sequence generation task, assigning each document a document identifier (DocID) and retrieving it through autoregressive decoding, making DocID design a critical factor in retrieval quality. However, existing schemes based on discrete representation learning suffer from inherent collision issues and create a mismatch between the DocID's encoding objective and the system's business optimization target. To address these limitations, we propose \textbf{Cluster-Ranked Identifier (CRID)}, which decouples DocID into \textit{semantic clustering} and \textit{business-value ranking}, yielding collision-free identifiers that support incremental updates via intra-cluster reranking. We further introduce an analytical framework that decomposes retrieval gains into \textit{personalized preference} and \textit{statistical prior} generalization, revealing how semantic cluster size governs the balance between the two components. Experiments on a Taobao e-commerce corpus of over 300M items show that CRID surpasses the strongest embedding-based retrieval baseline on top-K Hitrate, and delivers +1.06\% GMV in full-traffic deployment.
\end{abstract}

\section{Introduction}

\begin{figure*}[!t]
    \centering
    \includegraphics[width=1\linewidth]{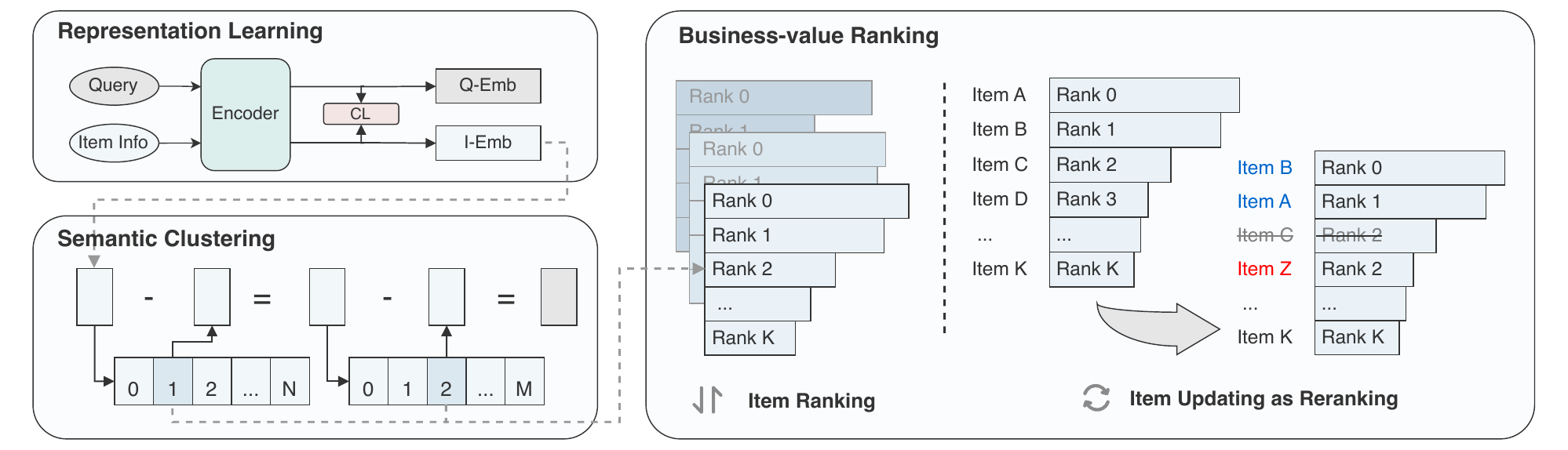}
    \caption{Overview of CRID. Left: item embeddings are learned via contrastive learning on query-item pairs (top) and quantized into a two-level semantic codebook (bottom). Right: items within each semantic cluster are ranked by business value to form the last-level DocID token (Item Ranking); incremental updates are performed by reranking items within affected clusters when new items arrive (Item Updating as Reranking).}
    \label{fig:main}
\end{figure*}

Generative Retrieval (GR) has recently emerged as a promising paradigm for information retrieval tasks such as e-commerce search and content search~\cite{chen2025onesearch, pang2025generative}. It assigns each document a document identifier (DocID) and retrieves it through autoregressive decoding, making DocID design a fundamental determinant of retrieval performance.

Existing DocID schemes span diverse categories including hierarchical IDs~\cite{rajput2023recommender, chen2025onesearch}, parallel IDs~\cite{xing2025reg4rec, xu2025store}, and other surrogate representations such as text terms~\cite{zhang2025c2t}. Among them, the coarse-to-fine hierarchical structure is highly compatible with the autoregressive generation process of large language models (LLMs), making it the prevailing choice in industrial systems. Most hierarchical schemes rely on discrete representation learning, such as RQ-KMeans~\cite{luo2025qarm} and RQVAE~\cite{lee2022autoregressive}, which discretize semantic embeddings into semantic IDs. Such methods suffer from an inherent collision problem, where multiple items may be assigned the same DocID, which becomes especially severe in large-scale corpora. To mitigate collisions, random IDs~\cite{fu2025forge} and balancing-based heuristics~\cite{zheng2024adapting} have been proposed as post-hoc remedies (see Appendix~\ref{appendix:related_work} for a detailed discussion of related work).

Beyond collisions, a more fundamental challenge arises: existing DocID schemes are constructed purely from semantic embeddings, creating a mismatch between the DocID's encoding objective (semantic reconstruction) and the system's optimization target (business conversion). For example, two items in the same semantic cluster may differ substantially in business value, yet receive adjacent or even identical DocIDs under purely semantic quantization. In large-scale industrial scenarios such as Taobao e-commerce search, well-optimized embedding-based retrieval (EBR) methods naturally incorporate business signals such as click-through rate and conversion rate through feature-rich training, which are entirely absent from current DocIDs. The gap is amplified by capacity–corpus tension: the candidate pool contains hundreds of millions of items, but the GR model cannot scale arbitrarily due to latency constraints, making the information efficiency of each DocID token critical. This objective mismatch, not model capacity alone, is the primary bottleneck.

To address these challenges, we propose \textbf{Cluster-Ranked Identifier (CRID)}, a simple yet effective DocID design that decouples each identifier into \textit{semantic clustering} and \textit{business-value ranking}: the former captures coarse, query-level semantic similarity, while the latter encodes statistical priors as ordinal ranks within each cluster. This construction eliminates collisions and allows incremental updates via intra-cluster reranking. The rank axis thereby carries a transferable business-value prior that aligns the identifier with the optimization target, without sacrificing collision-freeness. We further reveal that CRID's gains arise from two complementary mechanisms: \textit{personalized preference} and \textit{statistical prior} generalization, and that semantic cluster size controls their balance.

Our contributions are twofold. \textbf{(1)} CRID is, to our knowledge, the first DocID scheme to encode business value as ordinal ranking within semantic clusters. This deliberately simple design change at the last codebook level yields larger gains than sophisticated quantization techniques, while producing collision-free identifiers that support incremental updates through intra-cluster reranking without codebook retraining. \textbf{(2)} We provide an analytical framework that decomposes retrieval gains into personalized preference and statistical prior generalization, explaining why business-value ranking is effective and how semantic cluster size governs the trade-off between the two components. While developed in the context of CRID, this framework generalizes to hierarchical DocID schemes beyond CRID and directly guided the codebook configuration selected for production deployment. 

Experiments on a Taobao corpus of over 300M items show that CRID surpasses all baselines including the strongest EBR system on top-K Hitrate, and online A/B testing on full production traffic confirms +1.06\% improvement in GMV.

\section{Method}

\begin{table*}[!t]
    \centering
    \small
    \caption{Comparison of DocID schemes on in-search and out-of-search conversion Hitrate. All methods share the same first two codebook levels ($8192 \times 8192$ RQ-KMeans semantic clustering); each row denotes a different strategy applied to the third codebook level. ``Collision-free'' indicates whether the scheme guarantees a one-to-one mapping between DocIDs and items. All Hitrate metrics are computed at the item level. Best results are in bold.}
    \begin{tabular}{lccccccccc}
    \hline
        ~ & \multirow{2}{*}{Collision-free} & \multicolumn{4}{l}{In-search conversion} & \multicolumn{4}{l}{Out-of-search conversion} \\
        ~ & ~ & HR20 & HR100 & HR500 & HR1000 & HR20 & HR100 & HR500 & HR1000 \\ \hline
        $8192^3$ & N & 16.68\% & 33.73\% & 55.01\% & 63.83\% & 13.18\% & 26.51\% & 43.16\% & 49.99\% \\
        w/ OPQ & N & 20.86\% & 39.02\% & 59.86\% & 67.54\% & 16.32\% & 30.51\% & 47.10\% & 53.55\% \\
        w/ SK & N & 24.45\% & 44.00\% & 63.16\% & 69.90\% & 19.32\% & 34.26\% & 49.79\% & 55.46\% \\
        w/ Tiger & Y & 37.48\% & 51.83\% & 66.44\% & 73.15\% & 30.00\% & 39.96\% & 52.66\% & 57.99\% \\
        w/ FORGE & N & 37.28\% & 51.60\% & 66.29\% & 72.50\% & 29.66\% & 39.99\% & 51.99\% & 57.40\% \\
        w/ CRID & Y & \textbf{41.20\%} & \textbf{59.02\%} & \textbf{76.22\%} & \textbf{82.25\%} & \textbf{32.43\%} & \textbf{45.71\%} & \textbf{59.79\%} & \textbf{65.50\%} \\ \hline
    \end{tabular}
    \label{table:main}
\end{table*}

\paragraph{Problem Formulation.}
Given a user query $q$ and historical behavior sequence $\mathbf{h} = (h_1, h_2, \ldots, h_T)$, GR formulates retrieval as an autoregressive generation task. Each item is assigned a DocID $\mathbf{c} = (c_1, c_2, \ldots, c_L)$, where $L$ denotes the number of codebook levels. The GR model parameterized by $\theta$ generates the DocID tokens autoregressively:
\begin{equation}
    P_\theta(\mathbf{c} \mid q, \mathbf{h}) = \prod_{\ell=1}^{L} P_\theta(c_\ell \mid c_{<\ell}, q, \mathbf{h}).
\end{equation}
During inference, constrained beam search is used to decode a ranked list of candidate DocIDs according to $P_\theta(\mathbf{c} \mid q, \mathbf{h})$.

\subsection{Cluster-Ranked Identifier}

The overall architecture of CRID is illustrated in Fig.~\ref{fig:main}. A key observation is that in hierarchical DocIDs, earlier levels predominantly determine semantic recall while later levels refine disambiguation within the recalled set. To address the objective mismatch between semantic DocIDs and business optimization targets, we decompose the DocID into a semantic cluster prefix $\mathbf{c}_s = (c_1, \ldots, c_{L-1})$ and a business-value rank $r$ within that cluster. For a candidate item $i$, the generation probability can thus be factored as:
\begin{equation}
    P_\theta(i \mid q, \mathbf{h}) = P_\theta(\mathbf{c}_s \mid q, \mathbf{h}) \cdot P_\theta(r \mid \mathbf{c}_s, q, \mathbf{h}),
\end{equation}
where the first term captures semantic cluster selection and the second captures intra-cluster ranking.
By modeling semantics and business value jointly rather than applying value as a post-hoc rerank, the model learns the trade-off between them within one decoding objective, raising the ceiling that fine-ranking can reach under a fixed recall budget.

\begin{table}[!t]
    \centering
    \small
    \setlength{\tabcolsep}{3pt}
    \caption{Joint modeling (CRID) versus post-hoc reranking by business value on a two-level $8192^2$ semantic baseline. The rerank variants differ in the pool size. In-search conversion Hitrate (\%).}
    \begin{tabular}{lcccc}
    \hline
        ~ & HR20 & HR100 & HR500 & HR1000 \\ \hline
        + rerank (pool 10k) & 14.77\% & 33.73\% & 61.28\% & 70.97\% \\
        + rerank (pool 5k) & 17.54\% & 38.77\% & 62.45\% & 69.18\% \\
        + rerank (pool 1k) & 20.99\% & 37.42\% & 47.46\% & 48.62\% \\
        CRID & \textbf{41.20\%} & \textbf{59.02\%} & \textbf{76.22\%} & \textbf{82.25\%} \\ \hline
    \end{tabular}
    \label{table:inference_rerank}
\end{table}

\textbf{Semantic clustering.}
In search scenarios, user queries provide strong semantic anchors that naturally constrain the retrieval space. We train item embeddings via contrastive learning~\cite{oord2018representation} on query-item pairs, so that semantically relevant items are mapped closer together in the embedding space. After quantization with standard methods such as RQ-KMeans~\cite{luo2025qarm}, the resulting clusters capture query-level semantic granularity, sufficient for distinguishing relevance at the coarse level while leaving fine-grained discrimination to the business-value rank.

\textbf{Business-value ranking.}
While semantic clusters group items at query-level granularity, items within the same cluster can differ substantially in business value. To encode this variation, we rank items within each cluster by a business-value statistic (e.g., conversion rate) and use the rank as the last-level DocID token, with all items in the same cluster sharing a single rank codebook. This ordinal encoding naturally aligns the DocID structure with the GR model's optimization target, typically a business metric rather than a semantic category.

This design has two key advantages. \textbf{(1) Inherently collision-free}: each rank within a cluster maps to exactly one item, eliminating collisions by construction. \textbf{(2) Incremental updates}: newly added items are assigned to the nearest semantic cluster by embedding distance and reranked by updated business-value statistics on a daily schedule, requiring no codebook retraining. Unlike categorical attribute hashing or binning strategies~\cite{xue2026generative, zhang2026unified} that discretize business signals into unordered buckets, ordinal ranking arranges items along the rank axis by business value, so the last level carries a value-aligned statistical prior that transfers across clusters rather than an arbitrary identifier to memorize. We provide a deeper analysis of CRID's effectiveness in \S\ref{sec:analysis}.

\begin{figure}[!t]
    \centering
    \includegraphics[width=1\linewidth]{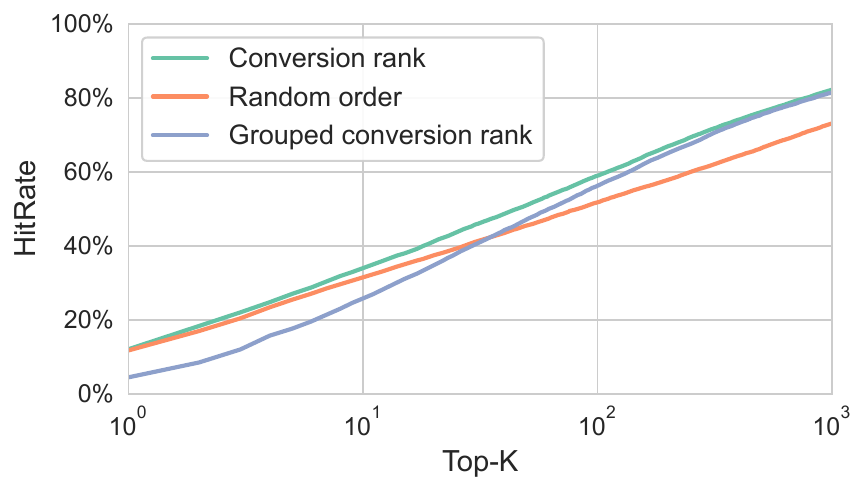}
    \caption{Ablation on business-value rank and collision-freeness. Conversion rank consistently outperforms Random order (no business-value signal) and Grouped conversion rank (grouping every four items).}
    \label{fig:fig1_business_rank_ablation}
\end{figure}


\section{Experiments}

\subsection{Experimental Setup}

We conduct all experiments on a candidate pool of over 300M items from the Taobao e-commerce search scenario, using Qwen2.5-0.5B~\cite{yang2024qwen2} as the backbone model. 
The first two codebook levels adopt an $8192 \times 8192$ semantic clustering configuration, with business-value rank as the third level. For evaluation, we report item-level Hitrate@$K$ (HR@$K$), measuring whether the ground-truth item appears in the top-$K$ predictions. Hitrate is the appropriate metric for the retrieval stage, as downstream fine-ranking determines the final presentation order. We distinguish \textbf{top-K Hitrate} (e.g., HR@20) from \textbf{deep-K Hitrate} (e.g., HR@1000). Evaluation samples include both in-search and out-of-search conversions, where the latter refers to conversions in the same product category as the query but outside the search scenario. Full details are provided in Appendix~\ref{appendix:setup}.

\begin{table}[!t]
    \centering
    \small
    \setlength{\tabcolsep}{2.5pt}
    \caption{Within-tier shuffle control. Item order is shuffled inside a window $W$ while coarse value tiers remain collision-free. $W\!=\!1$ is CRID; $W\!=\!$\,full removes all ordering. All variants share the same rank-token frequency. In-search HR@1000 (\%).}
    \begin{tabular}{lcccccc}
    \hline
        $W$ & 1 & 4 & 16 & 128 & 512 & full \\ \hline
        HR & \textbf{82.25\%} & 81.39\% & 81.67\% & 79.09\% & 76.29\% & 73.15\% \\
        $\Delta$ & -- & $-$0.86 & $-$0.58 & $-$3.16 & $-$5.96 & $-$9.10 \\ \hline
    \end{tabular}
    \label{table:tier_shuffle}
\end{table}

\subsection{Main Results and Ablations}
\label{sec:main_results}

\textbf{Main Results.} Starting from a base $8192^3$ RQ-KMeans codebook, we evaluate several strategies that modify the third codebook level including Sinkhorn-Knopp (SK) balancing~\cite{cuturi2013sinkhorn, zheng2024adapting}, OPQ discretization~\cite{ge2013optimized, chen2025onesearch}, and two random ID strategies: Tiger~\cite{rajput2023recommender}, which assigns monotonically increasing IDs for strict collision-freeness, and FORGE~\cite{fu2025forge}, which allows up to five items per DocID. CRID replaces the third-level codebook with business-value rank based on 30-day conversion counts. As shown in Table~\ref{table:main}, CRID outperforms all baselines on both in-search and out-of-search conversion Hitrate. Compared with the strongest baseline, CRID improves HR@20 by 3.72\,pp and HR@1000 by 9.10\,pp on in-search conversion.

For conciseness, we present ablation results on in-search conversion only; the conclusions for out-of-search conversion are consistent.

\textbf{Joint modeling outperforms post-hoc reranking.}
We compare encoding the business signal through joint modeling (CRID) against applying it as a post-hoc rerank. For the rerank baseline, a two-level $8192^2$ semantic model retrieves the top-1000 semantic clusters, expands them to member items, forms a top-$N$ rerank pool by decoding score, and reranks by the same global business-value statistic. Table~\ref{table:inference_rerank} shows that joint modeling dominates post-hoc reranking at every cutoff ($\sim$20\,pp at HR@20 and $\sim$11\,pp at HR@1000). The failure is structural: a query-agnostic rerank floods the head with popular-but-irrelevant items, while a smaller pool improves top-$K$ only by capping deep-$K$ recall. Business value must therefore be modeled jointly with the semantic prefix.

\textbf{Business value and collision-freeness are complementary.}
As shown in Fig.~\ref{fig:fig1_business_rank_ablation}, we replaced Conversion rank with Random order and Grouped conversion rank, where the latter groups every four consecutive items under one DocID to simulate mild collisions.
Random order leads to a substantial Hitrate drop at deep-K, while Grouped conversion rank degrades performance more noticeably at top-K. These results confirm that both properties are essential and complementary: business value primarily benefits deep-K Hitrate, while collision-freeness is more critical for top-K performance.

\textbf{Business-value ordering underlies the gain.}
CRID, Tiger, and FORGE share the same frequency prior on the last-level rank token, where low ranks are common and high ranks are rare, so this prior alone cannot explain CRID's advantage. What matters is aligning business value with it. We keep collision-free coarse value tiers and shuffle item order within windows of size $W$, from $W=1$ (CRID) to full-cluster random. Table~\ref{table:tier_shuffle} shows that all settings share the same rank-token frequency, yet HR@1000 spans 9.10\,pp. Shuffling within small windows ($W \leq 16$) costs under 0.9\,pp, so the effect is not tied to exact positions. Performance degrades only when ordering across materially different value levels is destroyed. Thus CRID exploits a coarse business-value ordering that is robust to fine-grained reordering.

\begin{table}[!t]
    \centering
    \small
    \caption{Effect of different business-value ranking signals on conversion Hitrate (top), and pairwise Spearman's $\rho$ among the three ranking orders (bottom). $\rho$ is computed over clusters with more than one item.}
    \begin{tabular}{lcccc}
    \hline
        ~ & HR20 & HR100 & HR500 & HR1000 \\ \hline
        Conv.-rank & \textbf{41.20\%} & 59.02\% & \textbf{76.22\%} & \textbf{82.25\%} \\
        Click-rank & 40.80\% & 58.30\% & 75.73\% & 81.81\% \\
        Score-rank & 41.15\% & \textbf{59.34\%} & 76.21\% & 81.82\% \\ \hline
    \end{tabular}

    \vspace{2mm}
    \begin{tabular}{lccc}
    \hline
        ~ & Conv.--Click & Conv.--Score & Click--Score \\ \hline
        $\rho$ & 0.708 & 0.666 & 0.617 \\
    \hline
    \end{tabular}
    \label{table:business_rank_ablation}
\end{table}

\begin{table*}[!t]
    \centering
    \small
    \caption{Incremental update evaluation 10 days after the training cutoff.}
    \begin{tabular}{lcccccc}
    \hline
        ~ & Pool Coverage & Avg. Items/CRID & HR20 & HR100 & HR500 & HR1000 \\ \hline
        No update & 65.08\% & \textbf{1.00} & 37.86\% & 54.40\% & 71.60\% & 77.52\% \\
        Insert only & 99.15\% & 6.31 & 29.13\% & 46.00\% & 63.83\% & 70.44\% \\
        Full rerank & \textbf{99.37\%} & \textbf{1.00} & \textbf{39.12\%} & \textbf{56.74\%} & \textbf{74.04\%} & \textbf{80.05\%} \\ \hline
    \end{tabular}
    \label{table:incremental_hitrate}
\end{table*}

\textbf{Robust to business-value definition.}
We further compare three business-value ranking signals: Conversion rank and Click rank, both computed from 30-day cumulative statistics, and a quality score predicted by an internal business model. Table~\ref{table:business_rank_ablation} reports Hitrate under each signal along with pairwise Spearman's $\rho$. Spearman's $\rho$ measures the rank-order agreement between two business-value rankings of items within the same semantic cluster. Despite moderate rank correlation ($\rho \approx 0.6$--$0.7$), downstream Hitrate differences remain limited, indicating that the GR model is robust to the choice of ranking signal.

\subsection{Incremental Update Experiments}

To evaluate CRID's incremental update capability, we test on data collected 10 days after the training cutoff (Table~\ref{table:incremental_hitrate}). We compare three settings: \textit{No update} (frozen codebook), \textit{Insert only} (newly arrived items are assigned to the nearest semantic cluster and mapped to the existing rank position with the closest business value), and \textit{Full rerank} (new and existing items jointly reordered by daily business-value statistics).

The \textit{Insert only} strategy achieves near-complete pool coverage (99.15\%) but degrades Hitrate substantially, as each CRID contains an average of 6.31 items, effectively reintroducing collisions. In contrast, \textit{Full rerank} maintains collision-free mapping while achieving the highest coverage (99.37\%) and the best Hitrate across all metrics, even surpassing the \textit{No update} setting. The latter's frozen codebook suffers from progressively invalidated DocID paths as items expire from the pool, effectively reducing its retrievable candidate set.

\begin{table}[!t]
    \centering
    \small
    \setlength{\tabcolsep}{4pt}
    \caption{Hitrate gains of the fully-trained CRID-based GR model over the strongest personalized EBR baseline.}
    \label{table:gr_vs_dr_hitrate}
    \begin{tabular}{lcccc}
    \hline
        ~ & HR20 & HR100 & HR500 & HR1000 \\ \hline
        In-search & \textbf{+13.26\%} & \textbf{+10.05\%} & -0.63\% & -3.02\% \\
        Out-of-search & \textbf{+8.00\%} & \textbf{+7.01\%} & \textbf{+3.75\%} & \textbf{+2.66\%} \\ \hline
    \end{tabular}
\end{table}

\begin{figure*}[!t]
    \centering
    \includegraphics[width=1\linewidth]{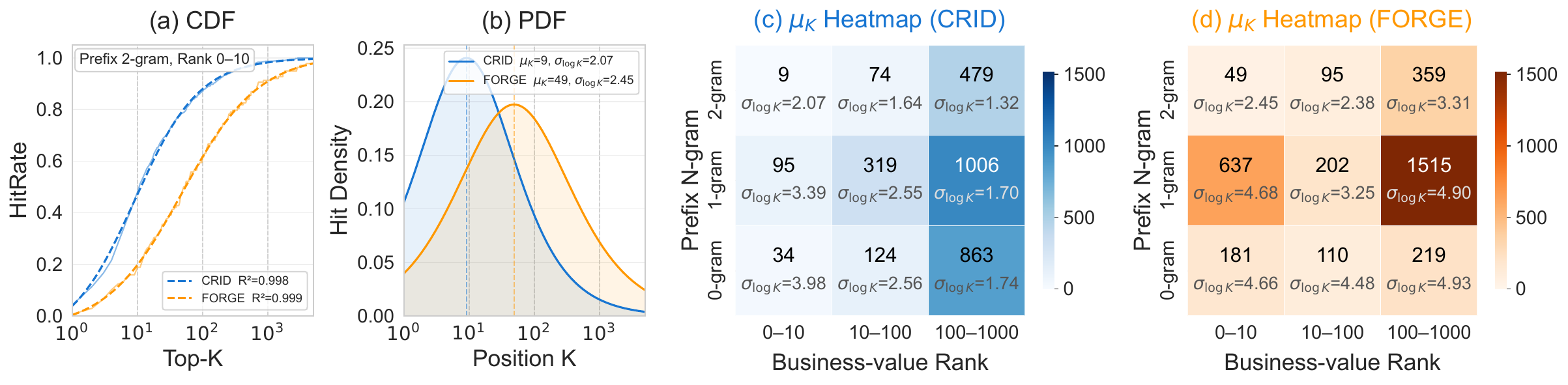}
    \caption{Logistic CDF fit and hit-density (PDF) analysis of CRID vs.\ FORGE. (a) Logistic fit to Hitrate CDF; (b) derived hit-density distribution 
    with fitted $\mu(K)$ and $\sigma(\log K)$; both for the Prefix 2-gram, Rank 0--10 group. (c)--(d) Heatmaps of fitted $\mu(K)$ across all prefix N-gram $\times$ rank groups, with per-cell $\sigma(\log K)$. Prefix 3-gram and rank 1000+ excluded (near-saturated / small sample). Full per-group curves in Appendix~\ref{appendix:hit_density}.}
    \label{fig:fig3g_heatmap_dual_annot}
\end{figure*}

\begin{figure}[!t]
    \centering
    \includegraphics[width=0.95\linewidth]{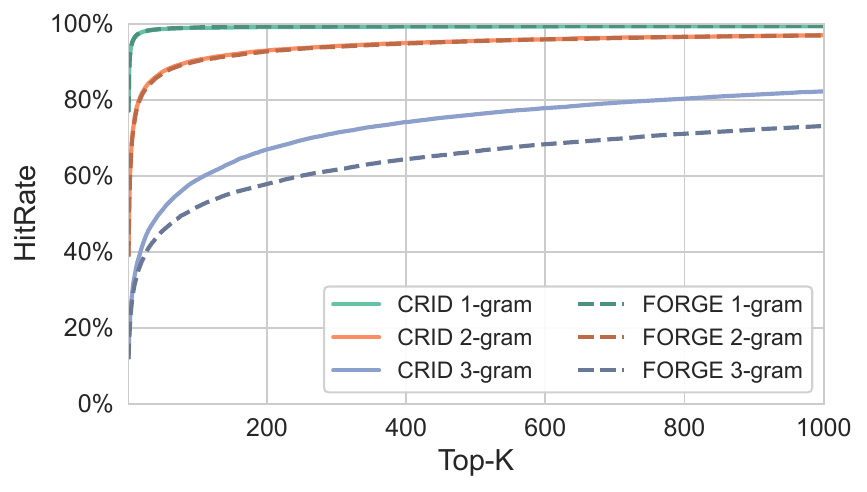}
    \caption{Prefix N-gram Hitrate@$K$ for CRID and the FORGE baseline. The 1-gram and 2-gram curves of the two methods nearly overlap
    . The gap appears entirely at the 3-gram (full-path) level.}
    \label{fig:fig2_prefix_n_gram_hitrate}
\end{figure}

\subsection{Full-Scale Evaluation and Deployment}
\label{sec:fullscale}

\textbf{Offline comparison with EBR.}
For deployment, we train CRID with a production-grade pipeline using a $32768 \times 8192$ semantic codebook with a business-value rank range of $8192$, configured via the analysis in \S\ref{sec:analysis} (setup details are provided in Appendix~\ref{appendix:setup}). 
The deployed $32768 \times 8192$ codebook yields realized clusters with a mean size of 8.75.
Table~\ref{table:gr_vs_dr_hitrate} compares the fully-trained CRID-based GR model against the strongest personalized EBR baseline. For in-search conversion, CRID achieves substantial top-K gains (+13.26\% HR@20) with slight degradation at deeper cutoffs. Out-of-search gains are more pronounced and consistent across cutoffs (+8.00\% HR@20, +2.66\% HR@1000), indicating that CRID provides complementary recall beyond existing retrieval pipelines and generalizes better than EBR.

\textbf{Online deployment.}
The GR model is deployed as an additional retrieval channel alongside existing EBR pipelines, operating as a unified retrieval and coarse-ranking pipeline. It uses constrained decoding with dynamic beam sizes of 100, 400, and 1500 across the three decoding stages, determined via cumulative probability cutoffs to ensure near-lossless Hitrate coverage (details in Appendix~\ref{appendix:beam_size}). The GR model recalls approximately 1300 items into the fine-ranking stage, meeting the same latency requirements as existing retrieval channels. A 1\% traffic A/B test over 30 days yielded +0.18\% IPV (item page views), +0.54\% order count, and +1.06\% GMV (Gross Merchandise Value, the total monetary value of transactions completed through the platform) for overall traffic. The method has since been deployed to full production traffic.

\section{Analysis}
\label{sec:analysis}

To understand the source of CRID's gains and offer generalizable analytical insights for hierarchical DocID design, we attribute retrieval improvements to two complementary mechanisms: \textit{personalized preference generalization}, the model's ability to leverage sequential patterns in user history, and \textit{statistical prior generalization}, its ability to generalize from corpus-level business-value statistics. To validate and quantify these two components, we introduce an analysis framework based on prefix N-gram and business-value rank stratification. We first verify that the statistical prior encoded by business-value rank is effective (\S\ref{sec:why_bvr}), then quantify the respective contributions of the two components (\S\ref{sec:gain_decomposition}), and finally examine how semantic cluster size affects the net retrieval gain (\S\ref{sec:cluster_size}).

\subsection{Effectiveness of Statistical Prior}
\label{sec:why_bvr}

We evaluate prediction accuracy at each codebook level using \textit{Prefix N-gram Hitrate@$K$}, defined as whether the \emph{predicted} prefix N-gram matches that of the \emph{ground-truth label}. We retain one label per evaluation sample to avoid duplicate matches. Prefix N-gram Hitrate is computed at the DocID level: a prediction is a hit if its DocID prefix matches the ground-truth's, regardless of whether the full DocID resolves to the correct item.

As shown in Fig.~\ref{fig:fig2_prefix_n_gram_hitrate}, both CRID and the FORGE baseline ($8192^3$) achieve comparable 1-gram and 2-gram Hitrate, confirming that the first two semantic levels are equally effective. However, CRID's 3-gram (full-path) Hitrate significantly exceeds FORGE's (82\% vs 72\% at $K$=1000). This gap indicates that statistical prior generalization alleviates the last-level bottleneck, yielding larger gains than fine-grained semantic discrimination.

\begin{figure*}[!t]
    \centering
    \includegraphics[width=1\linewidth]{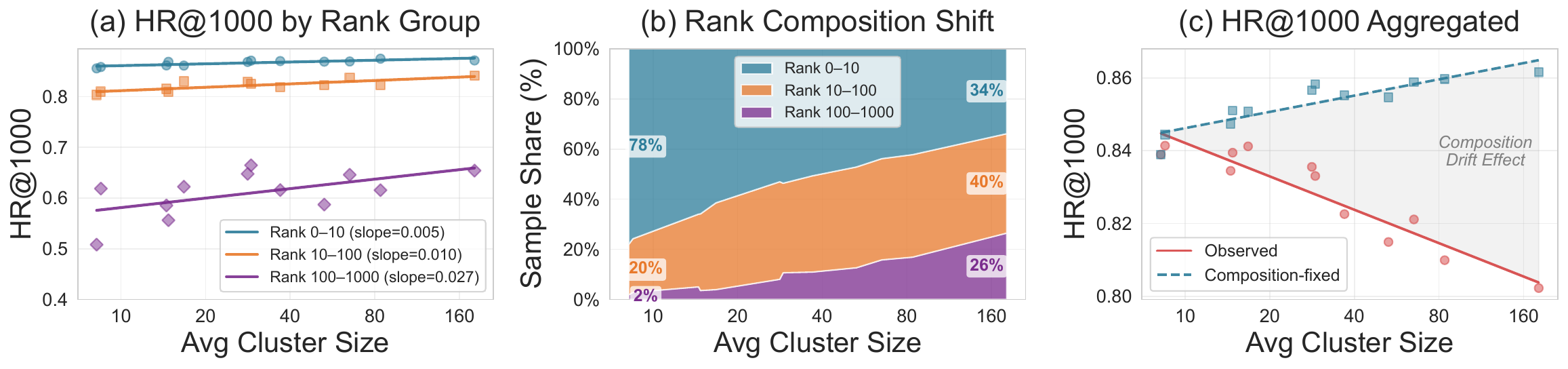}
    \caption{Composition effect on HR@1000 across codebook configurations with varying average semantic cluster sizes (rank perspective). (a) Per-rank-group HR@1000 increases with cluster size. (b) As cluster size increases, the proportion of top-ranked items (Rank 0--10) shrinks from 78\% to 34\%. (c) This composition shift causes the weighted HR@1000 to decrease despite per-group improvements. See Appendix~\ref{appendix:composition} for details.}
    \label{fig:fig4_cluster_composition}
\end{figure*}

\subsection{Gain Decomposition}
\label{sec:gain_decomposition}

We operationalize the two components by partitioning the evaluation set along two dimensions. For personalized preference, we group samples by whether the target item's DocID prefix appears in the user's historical behavior sequence, yielding four groups: \textit{prefix 3-gram} (full-path match), \textit{2-gram}, \textit{1-gram}, and \textit{0-gram} (no match). This measures behavioral overlap, distinct from Prefix N-gram Hitrate in \S\ref{sec:why_bvr}. For statistical prior, we divide items into four business-value rank groups: \textit{rank 0--10}, \textit{10--100}, \textit{100--1000}, and \textit{1000+}.

For systematic comparison, we fit each group's Hitrate curve as a logistic CDF on $\log K$ (Fig.~\ref{fig:fig3g_heatmap_dual_annot}(a)) and differentiate to obtain the \emph{hit-density distribution} (Fig.~\ref{fig:fig3g_heatmap_dual_annot}(b)), a PDF characterizing how hits spread across recall depths (full per-group fits in Appendix~\ref{appendix:hit_density}, with coefficient of determination $R^2 > 0.99$ for most groups). Its mean $\mu(K)$ indicates the \emph{sweet spot}, the recall depth at which a group is best supported (smaller is better), and $\sigma(\log K)$ captures the dispersion around it.

As shown in Fig.~\ref{fig:fig3g_heatmap_dual_annot}(c)(d), CRID achieves substantially lower $\mu(K)$ than FORGE for top-ranked items (Rank 0--10) and uniformly smaller $\sigma(\log K)$ across all groups, confirming that business-value ranking concentrates hits into shallower recall depths with tighter concentration. Within CRID, $\mu(K)$ increases monotonically along the rank axis, a structured pattern absent in FORGE, indicating that ordinal ranking establishes an ordered retrieval structure across the rank space. This concentration of gains at shallow ranks directly improves overall Hitrate.

\subsection{Impact of Semantic Cluster Size}
\label{sec:cluster_size}

Prior studies have shown that codebook configuration affects retrieval performance~\cite{ju2025generative}, but provide limited guidance on how to choose it for a given corpus. A naive expectation is that larger clusters, which strengthen statistical prior by providing more items to rank, should uniformly improve performance. We show that the actual effect is more nuanced due to a composition shift that opposes per-group gains. We use \textit{average semantic cluster size}, defined as the total number of items divided by the number of semantic clusters, as the metric to analyze how cluster granularity affects retrieval. We evaluate a range of codebook configurations, including both uniform (e.g., $8192^2$) and non-uniform (e.g., $65536 \times 1024$) semantic configurations, with the business-value rank level fixed at $8192$, sufficient for all configurations tested. This yields cluster sizes spanning roughly 5 to 160. Full per-group curves are in Appendix~\ref{appendix:composition}.

As shown in Fig.~\ref{fig:fig4_cluster_composition}, from the business-value rank perspective, per-rank-group HR@1000 increases with cluster size across all rank ranges (panel a), yet the actual weighted average decreases (panel c). This reversal is driven by composition shift (panel b): as cluster size increases, the proportion of top-ranked items (Rank 0--10) shrinks from 78\% to 34\%, diluting the overall Hitrate despite per-group improvements. The prefix N-gram perspective exhibits a similar pattern (see Appendix~\ref{appendix:composition} for full analysis), where smaller clusters improve per-group Hitrate but shift samples toward weaker prefix-match groups, limiting overall gains. The net effect of cluster size thus reflects opposing forces across the two dimensions: from the rank perspective, smaller clusters shift composition toward top-ranked groups but weaken per-group statistical prior; from the prefix perspective, smaller clusters improve per-group Hitrate but shift composition toward weaker prefix-match groups, precluding a universally optimal cluster size.

\subsection{Prerequisites and Practical Guidance}

These findings rely on two prerequisites. First, the optimization objective must correlate with the business-value signal used for ranking. This requirement is mild rather than brittle: any proxy correlated with the objective, such as conversion, click, or a model-predicted quality score, works comparably (Table~\ref{table:business_rank_ablation}), so the design does not hinge on a single hand-picked metric. Search naturally satisfies this, since conversion and GMV are well defined and aligned with the optimization target. Second, semantic clustering at the first levels must be sufficiently concentrated to sustain high prefix-level Hitrate. This too is satisfied in search scenarios, where queries provide strong semantic constraints on the retrieval space. In practice, we applied this framework to select the configuration targeting optimal HR@1000, directly informing the $32768 \times 8192$ production codebook (\S\ref{sec:fullscale}).

\section{Conclusion}

We presented Cluster-Ranked Identifier (CRID), a DocID design that decouples semantic clustering from business-value ranking to address the objective mismatch between semantic DocIDs and business optimization targets. By encoding business value as ordinal ranks within semantic clusters, CRID yields collision-free identifiers that support incremental updates through intra-cluster reranking. We additionally proposed an analysis framework that decomposes retrieval gains into personalized preference and statistical prior generalization, revealing how semantic cluster size governs the trade-off between them. Experiments on Taobao search demonstrate substantial top-K Hitrate improvements, and full-traffic deployment delivers +1.06\% GMV improvement. This framework provides actionable guidance for designing hierarchical DocIDs in large-scale retrieval systems.

\section*{Limitations}

Our experiments are conducted exclusively on Taobao e-commerce search. While CRID's design is not inherently search-specific, its effectiveness in other domains such as recommendation, where query semantics play a less central role, remains to be validated.

The business-value rank relies on historical statistics, introducing a potential cold-start bias: newly listed items with limited interaction data may receive suboptimal rank assignments, reducing their chance of being retrieved until sufficient statistics accumulate. Additionally, our design uses a single business-value signal; how to fuse multiple signals (e.g., conversion rate, click-through rate, and predicted scores) into a unified ordinal ranking remains an open direction.

Finally, all experiments use a 0.5B-parameter model due to online latency constraints. Larger models may alter the capacity–corpus balance that motivates CRID's design; exploring CRID's effectiveness under scaled-up model capacities is left for future work.

\bibliography{custom}

\clearpage
\appendix

\section{Related Work}
\label{appendix:related_work}

\subsection{Generative Retrieval}

Generative retrieval (GR) reframes retrieval as sequence-to-sequence generation, where a model directly produces document identifiers given an input query.
GENRE~\cite{de2020autoregressive} first demonstrated autoregressive retrieval by generating entity names with trie-constrained decoding.
DSI~\cite{tay2022transformer} formalized the paradigm by encoding corpus information within a single Transformer, mapping queries directly to document IDs.
NCI~\cite{wang2022neural} improved decoding with prefix-aware adaptation.
TIGER~\cite{rajput2023recommender} first introduced GR to recommendation, creating Semantic IDs via RQ-VAE and training a seq2seq model to predict next-item identifiers.
Subsequent works refer to such identifiers as \textit{Semantic IDs} or \textit{DocIDs}; we use the latter.
This paradigm has since been deployed at industrial scale: OneRec~\cite{zhou2025onerec} for video recommendation, OneSearch~\cite{chen2025onesearch} for e-commerce search, and GR4AD~\cite{xue2026generative}/GPR~\cite{zhang2025gpr} for advertising, demonstrating significant efficiency gains and business metric improvements.
Meanwhile, recent work improved SID-based GR training efficiency via token pruning and multi-step prediction~\cite{zhan2026semantic}.

\subsection{DocID Construction}

DocID design determines how items are discretized into the token space that generative models must learn to produce.

\textbf{Quantization-based methods.}
DSI~\cite{tay2022transformer} originally used hierarchical k-means clustering. TIGER~\cite{rajput2023recommender} introduced RQ-VAE with random suffixes for collision avoidance. LC-Rec~\cite{zheng2024adapting} applied Sinkhorn balancing to enforce uniform codebook utilization. Industrial systems further showed RQ-Kmeans offers superior balance: OneRec~\cite{zhou2025onerec} adopted balanced RQ-Kmeans, OneSearch~\cite{chen2025onesearch} employed OPQ to reduce information loss, and GPR~\cite{zhang2025gpr} proposed RQ-Kmeans+ (RQ-Kmeans initialization for RQ-VAE) to stabilize training. SA$^2$CRQ~\cite{wang2026towards} further challenged the fixed-length assumption by dynamically allocating variable-length codes based on item popularity. Beyond cascaded approaches, REG4Rec~\cite{xing2025reg4rec} and STORE~\cite{xu2025store} adopted parallel multi-expert quantization with orthogonal penalties to decompose high-cardinality features into compact token sets.

\textbf{Joint optimization and value alignment.}
DGI~\cite{wang2026differentiable} proposed differentiable geometric indexing to bridge the gap between DocID construction and downstream retrieval, though at the cost of increased coupling and deployment complexity.
In the advertising domain, GR4AD~\cite{xue2026generative} replaced the final quantization level with a hash-based mapping over categorical business attributes (e.g., conversion type, account ID), reducing the collision rate from 85\% to 18\% but not eliminating collisions. UniVA~\cite{zhang2026unified} adopted a classify-then-bin strategy that groups items by commercial attributes and discretizes bid values within each group. Although both approaches encode business signals at the last DocID level, they remain fundamentally fine-grained clustering or hash-based mappings that discretize business attributes into categorical buckets, without introducing the notion of ordinal ranking among items. As a result, they cannot guarantee collision-free assignment and require reconstruction when attribute definitions or value distributions change. In contrast, CRID uses simple ordinal ranking by a single business-value statistic, explicitly encoding a numerical order that the autoregressive model can generalize over, achieving collision-free assignment by construction and supporting incremental updates through intra-cluster reranking alone.

Subsequent work extends the value-ranking idea from item-level statistics to query-item cross statistics, enabling finer-grained personalization~\cite{zhan2026tsgr}.

\section{Experimental Details}
\label{appendix:setup}

We constructed CRID based on a candidate pool of over 300M items from the Taobao e-commerce search scenario. The corpus is assembled from Taobao search logs and the platform's item catalog and is proprietary. No public dataset citation is provided. For representation learning, we collected approximately 100M query-item pairs filtered by a relevance model and trained a shared-encoder model to produce 256-dimensional item embeddings via contrastive learning with in-batch negatives on 128 GPUs with a per-GPU batch size of 256. Since the CRID scheme and the discretization scheme are fully decoupled, we used RQ-KMeans to build semantic clusters. Specifically, we sampled 100M examples and applied hierarchical mini-batch KMeans with a batch size of 4M for 3 epochs. Unless otherwise specified, no auxiliary techniques such as balancing were used. Following mainstream codebook settings, we used an $8192 \times 8192$ configuration for the first two semantic codebook levels, and used business-value rank as the third level, with the rank space determined by the number of items in each cluster. We use a three-level codebook to balance discretization error and inference latency. CRID's collision-freeness holds at any depth.

The GR training pipeline consists of two stages: DocID memorization and Supervised Fine-Tuning (SFT). In the DocID memorization stage, the model is trained to predict DocID sequences from item features, warming up the newly added DocID vocabulary. To reduce training cost, we evaluated CRID using a 40M-sample subset of the training data. Empirically, models trained at this scale had largely converged, and the resulting performance was stable enough to reflect relative differences among methods. We did not observe any ranking inversion when training was extended further. Due to practical constraints such as online latency, all experiments were conducted using Qwen2.5-0.5B~\cite{yang2024qwen2}.

For full-scale online deployment (\S\ref{sec:fullscale}), we adopted a production-grade training pipeline with a larger-scale SFT stage and an additional Direct Preference Optimization (DPO)~\cite{rafailov2023direct} stage. Specifically, the SFT stage uses over 150M training samples, and the DPO stage samples 1M queries with an average of 10 response pairs per query for preference optimization. The DPO stage provides an incremental improvement of approximately 0.5\,pp in offline in-search conversion HR@1000. The CRID semantic codebook was configured as $32768 \times 8192$ with a business-value rank range of $8192$. The larger first-level vocabulary reduces average cluster size, favoring top-K performance as shown in \S\ref{sec:cluster_size}, while the rank range provides headroom for future pool expansion.

During inference, we used constrained beam search with a dynamic beam size. Specifically, we enforced valid-path constraints for the first two DocID positions, constraining the beam to only expand valid prefixes according to the codebook trie, with a beam size of 1000. For the last DocID position, we removed the trie constraint to reduce prefix-tree memory overhead and capped the beam size at 10,000, which was sufficient for the metrics considered in our experiments.

Out-of-search conversion samples refer to conversions where the purchased item belongs to the same product category as the query intent, but the conversion itself did not occur in the search scenario. Both conversion count and click count used for business-value ranking are computed over a 30-day rolling window.

\begin{figure}[!t]
    \centering
    \includegraphics[width=1\linewidth]{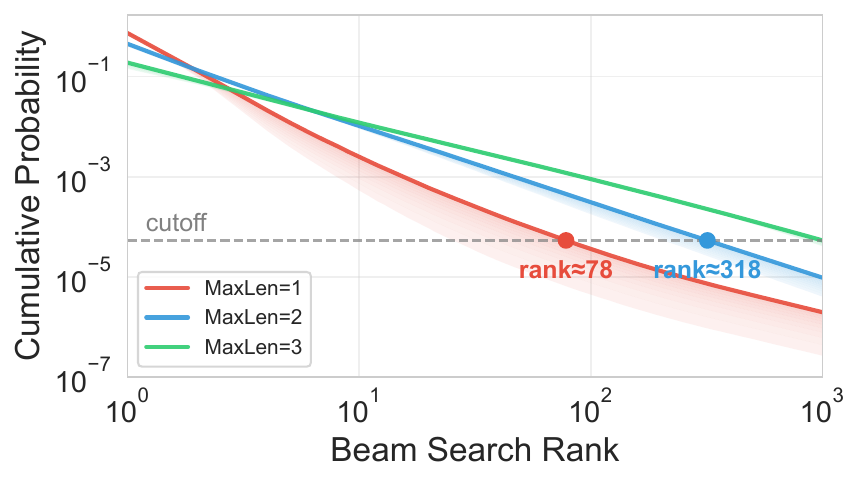}
    \caption{Cumulative probability of beam search outputs under different maximum new token settings (MaxLen = 1, 2, 3). Solid lines denote the mean probability; shaded regions illustrate the mean--median (P50) gap with gradient coloring fading toward the median.}
    \label{fig:fig6_response_probs_curve}
\end{figure}

\begin{figure*}[!t]
    \centering
    \includegraphics[width=1\linewidth]{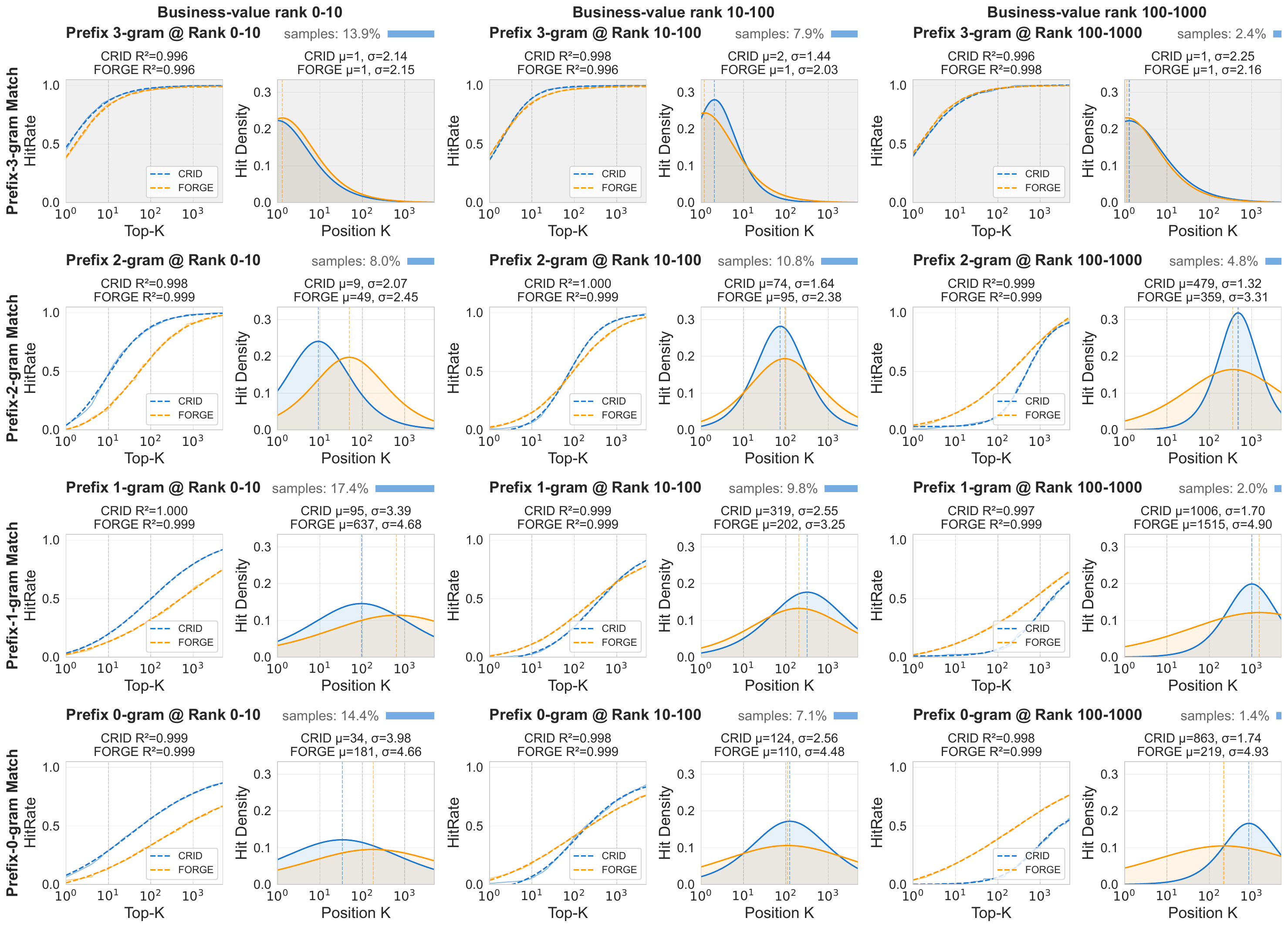}
    \caption{Per-group logistic CDF fits (left of each pair) and corresponding hit-density distributions (right) for CRID and FORGE across all prefix N-gram match groups (rows) and business-value rank groups (columns). Each panel is annotated with $R^2$, $\mu(K)$, $\sigma(\log K)$, and sample proportion. These are the full per-group fits summarized by the heatmap in Fig.~\ref{fig:fig3g_heatmap_dual_annot}.}
    \label{fig:fig3_full}
    \vspace{-0.2cm}
\end{figure*}

\section{Dynamic Beam Size Calibration}
\label{appendix:beam_size}

A three-level codebook requires stage-specific beam widths.
As shown in Fig.~\ref{fig:fig6_response_probs_curve}, the cumulative probability distribution over candidates becomes progressively flatter at later decoding stages: early stages concentrate probability mass on a small number of candidates, while later stages spread it across an increasingly long tail.
Applying a uniform beam size across all stages would over-allocate resources at early stages and under-allocate at later ones, wasting inference compute.

We calibrate stage-wise beam sizes by defining a probability cutoff at the MaxLen\,=\,3 curve at rank 1000 (dashed line in Fig.~\ref{fig:fig6_response_probs_curve}).
The ranks at which the MaxLen\,=\,1 and MaxLen\,=\,2 curves drop below this cutoff---approximately 78 and 318, respectively---indicate the effective beam width needed at each stage to maintain equivalent coverage.
This yields calibrated beam sizes of $\sim$80, $\sim$320, and 1000; the deployed setting of 100, 400, and 1500 (\S\ref{sec:fullscale}) rounds up to provide a safety margin.
The cutoff rank is configurable: adjusting it to match the desired recall budget yields the corresponding stage-wise beam sizes.

The shaded regions in Fig.~\ref{fig:fig6_response_probs_curve} reveal a notable mean--median gap, most pronounced at the first decoding stage (MaxLen\,=\,1) and diminishing at later stages. The gap indicates that query-level difficulty is most heterogeneous during semantic cluster selection: the majority of queries exhibit concentrated probability mass over a few top clusters, while a tail of ambiguous queries spreads probability across many candidates, pulling the mean above the median. As decoding progresses and the candidate space is narrowed by earlier stages, this difficulty disparity shrinks. The pronounced first-stage gap further motivates stage-specific beam sizing: relative to the median, a larger beam at the first stage is needed to accommodate the difficult queries whose probability mass is broadly dispersed.

\begin{figure*}[!t]
    \centering
    \renewcommand{\thesubfigure}{\Alph{subfigure}}
    \begin{subfigure}[t]{0.49\linewidth}
        \centering
        \includegraphics[width=\linewidth]{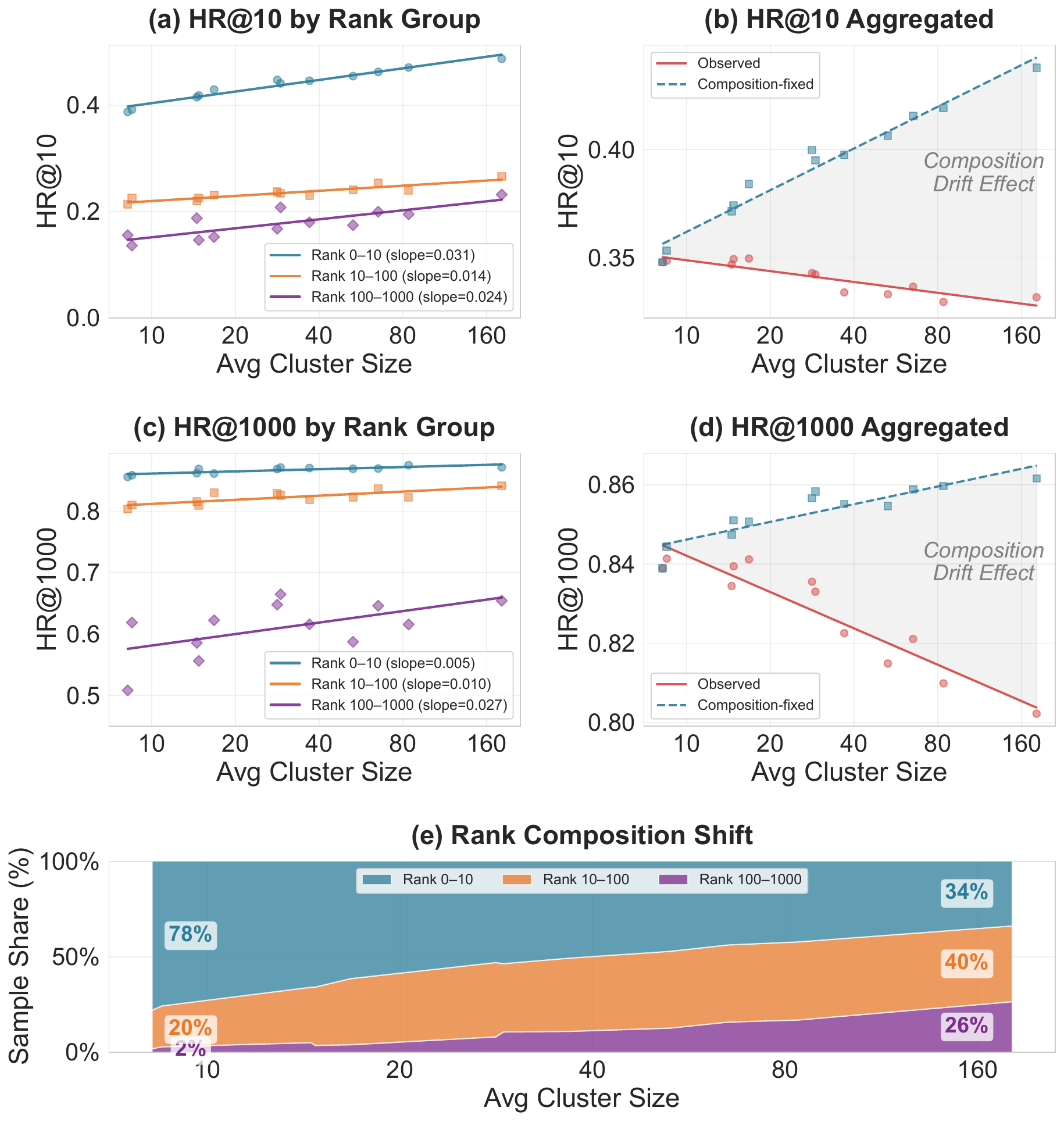}
        \caption{Business-value rank perspective.}
        \label{fig:fig4f_simpson_paradox_rank}
    \end{subfigure}
    \hfill
    \begin{subfigure}[t]{0.49\linewidth}
        \centering
        \includegraphics[width=\linewidth]{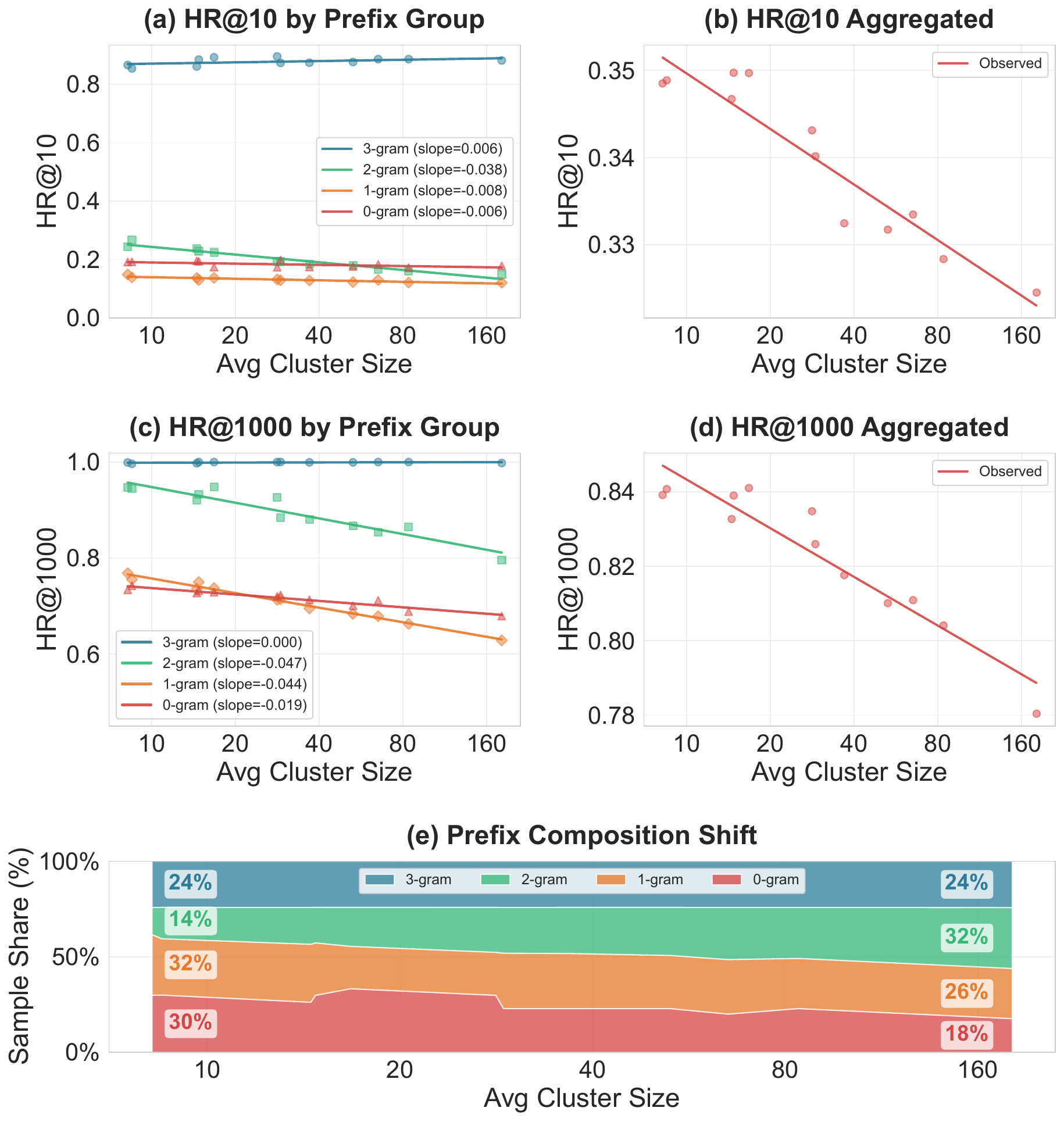}
        \caption{Prefix N-gram perspective.}
        \label{fig:fig4f_simpson_paradox_prefix}
    \end{subfigure}
    \caption{Full-resolution analysis of the composition effect across codebook configurations, extending Fig.~\ref{fig:fig4_cluster_composition} to both HR@10 and HR@1000. Each subfigure shows per-group Hitrate vs.\ average cluster size (left column), aggregated Hitrate trends (right column), and composition shift (bottom). The rank subfigure additionally overlays fixed-composition averages to isolate the composition drift effect.}
    \label{fig:fig4_full}
    \vspace{-0.2cm}
\end{figure*}

\section{Full Hit-Density Analysis}
\label{appendix:hit_density}

We treat each group's Hitrate on a log scale as a cumulative distribution function (CDF). Since Hitrate is monotonically increasing and bounded, we model it as a shifted and scaled CDF of a logistic distribution on $\log K$, with mean $\mu$ and scale $s$, plus offset and amplitude parameters to accommodate the bounded range of Hitrate. We then differentiate the fitted CDF to obtain the hit-density distribution, whose mean $\mu(K)$ and standard deviation $\sigma(\log K)$ characterize each group's retrieval behavior.

Fig.~\ref{fig:fig3_full} shows per-group logistic CDF fits and hit-density distributions for CRID and FORGE across all prefix N-gram $\times$ business-value rank groups, extending the heatmap summary in Fig.~\ref{fig:fig3g_heatmap_dual_annot}.

The \textit{prefix 3-gram} groups (top row) are excluded from the main-text heatmap because both methods achieve near-saturated Hitrate with nearly overlapping CDF curves, indicating negligible additional benefit from statistical prior generalization when a full-path prefix match exists. The \textit{rank 1000+} groups are excluded due to insufficient sample size for reliable fitting. For the remaining 3$\times$3 groups, the full per-group fits confirm the pattern discussed in \S\ref{sec:gain_decomposition}: CRID achieves substantially lower $\mu(K)$ than FORGE for top-ranked items and uniformly smaller $\sigma(\log K)$ across all groups, and its $\mu(K)$ increases monotonically along the rank axis, while FORGE exhibits no such structure.

In contrast, the \textit{prefix 0-gram} groups (bottom row) reveal that CRID retains a substantial advantage even without behavioral overlap: for Rank 0--10, CRID achieves $\mu(K) \approx 34$ versus FORGE's $\mu(K) \approx 181$, indicating that ordinal ranking provides strong statistical prior generalization independent of personalized preference.

A notable anomaly appears when comparing the prefix 1-gram and 0-gram groups: despite stronger behavioral overlap, the 1-gram group exhibits slightly higher $\mu(K)$ (i.e., worse performance) than the 0-gram group at small $K$. We attribute this to \emph{autoregressive prefix bias}: beam search assigns higher probability mass to candidate paths with longer prefix matches, disproportionately allocating beam slots to 2-gram and 3-gram candidates and ``crowding out'' 1-gram targets. The 0-gram group is less affected because it includes high-frequency, highly specific queries (e.g., ``iPhone 17 Pro Max'') whose target items are retrieved via semantic matching rather than sequential preference, reducing dependence on beam allocation among prefix-match groups.

\section{Full Composition Effect Analysis}
\label{appendix:composition}

Fig.~\ref{fig:fig4_full} extends the main-text analysis (Fig.~\ref{fig:fig4_cluster_composition}) to both HR@10 and HR@1000, from both the business-value rank and prefix N-gram perspectives.

\paragraph{Business-value rank perspective (Fig.~\ref{fig:fig4_full}A).}
The composition-driven reversal discussed in \S\ref{sec:cluster_size} is present at both HR@10 and HR@1000, but the effect is more pronounced at HR@10: per-group slopes are steeper (Rank 0--10: 0.031 for HR@10 vs.\ 0.005 for HR@1000), indicating that cluster size has a larger impact on top-K retrieval. Accordingly, the gap between the fixed-composition and actual weighted averages widens more rapidly at HR@10 (right-column panels), confirming that composition shift is the dominant factor limiting top-K performance under larger cluster sizes.

\paragraph{Prefix N-gram perspective (Fig.~\ref{fig:fig4_full}B).}
Unlike the rank perspective, per-group Hitrate does not uniformly increase with cluster size. The 3-gram group remains saturated ($\sim$1.0 for HR@1000, slope $\approx$ 0), while the 2-gram, 1-gram, and 0-gram groups all degrade as cluster size increases (HR@1000 slopes of $-$0.047, $-$0.044, and $-$0.019, respectively), reflecting that larger clusters dilute prefix coverage and increase intra-cluster disambiguation difficulty. Meanwhile, composition shifts toward stronger prefix groups as cluster size increases (2-gram share rises from 14\% to 32\%; 0-gram share drops from 30\% to 18\%), but this favorable shift cannot offset the per-group degradation, and the weighted average still declines. Thus, the prefix perspective also exhibits opposing forces: larger clusters shift composition toward stronger prefix-match groups but degrade per-group Hitrate. Combined with the rank perspective, where larger clusters improve per-group statistical prior but shift composition away from top-ranked groups, these opposing forces across the two dimensions preclude a universally optimal cluster size.

\end{document}